\documentstyle[eqsecnum,prd,aps]{revtex}
\makeatletter
\def\@maketitle{%
\vspace*{-2pc}
\vbox{\mbox{\large\sl\vbox to50pt{\hbox{Department of Physics}
               \hbox{Kyoto University}\vss}}
\hfill\mbox{\large\rm\vbox to50pt{\hbox{KUNS 1308}
                                  \hbox{December 1994}\vss}}
}\vspace{1pc}
\@title
\ifdim\prevdepth=-1000pt \prevdepth0pt\fi
\@authoraddress
\@date
}
\makeatother
\begin{document}
\draft
\title{Euclidean vacuum mode functions for a scalar field \\
on open de Sitter space}
\author{Misao Sasaki,\footnote{Address after January 1, 1995:
Department of Earth and Space Science, Osaka University, Toyonaka 560.
} Takahiro Tanaka and Kazuhiro Yamamoto}
\address{{\em Department of Physics,~Kyoto University}\\
{\em Kyoto 606-01,~Japan}}
\maketitle
\begin{abstract}
Motivated by recent studies of the one-bubble inflationary universe
scenario that predicts a low density, negative curvature universe,
we investigate the Euclidean vacuum mode functions of a scalar
field in a spatially open chart of de Sitter space which is foliated by
hyperbolic time slices.
When we consider the possibility of an open inflationary universe,
we are faced with the problem of the initial condition for
the quantum fluctuations of the inflaton field, because the
inflationary era should not last too long to lose every
information of the initial condition.
In the one-bubble scenario in which an open universe is created in
an exponentially expanding false vacuum universe triggered by
quantum decay of false vacuum, it seems natural that the initial
state is the de Sitter-invariant Euclidean vacuum.
Here we present explicit expressions
for the Euclidean vacuum mode functions in the open chart for a
scalar field with arbitrary mass and curvature coupling.
\end{abstract}

\section{Introduction}
Recently there appeared not a few observations which suggest
our universe has negative curvature, i.e., $\Omega_0\sim 0.1$
\cite{lowomega}. Accordingly, papers comparing theoretical
predictions of open universe models with observational data such as
COBE\cite{COBE} have been appearing in the context of inflationary
universe models\cite{openinf}, in cosmological models with topological
defects\cite{spergel} or in a general context by assuming a power-law type
primordial density perturbation spectrum\cite{sugiyama}.
However, in the standard inflationary universe paradigm, it is
generally believed that an open FRW universe with small perturbations
is hard to be realized in a consistent manner\cite{Kas}.

One possible consistent scenario is the creation of an open universe
from an exponentially expanding false vacuum dominated
universe\cite{styyPL,bucher,Bruce,sataya2}.
 Originally this idea was proposed by Gott\cite{Got}.
In the standard inflationary universe scenario,
the horizon problem is solved by a large amount of expansion of space.
A homogeneous patch initially of a horizon size expands
exponentially and our present horizon size will be inside such a
homogeneous patch.
However in this context, the flatness (or entropy) problem is solved
at the same time when the horizon problem is solved.
Therefore the spatial curvature at present time is inevitably
decreased by the expansion of universe.
This is the problem of the standard scenario of inflation
if we attempt to construct an open universe model
with $1-\Omega_0=O(1)$.
On the other hand, if we consider a bubble nucleation in the sea of
false vacuum which is described by de Sitter space,
the interior of a bubble has the $O(3,1)$ invariance owing to the
$O(4)$ symmetry of the Euclidean bounce solution which
represents the tunneling process\cite{CalCol}.
Thus the horizon problem is automatically solved.
Though the spacetime has the exact $O(3,1)$ invariance
in the lowest order description, quantum fluctuations around the
classical background may give rise to cosmological density
perturbations to explain the large scale structure of the universe.
However if the vacuum energy never becomes a dominant
component of the cosmic energy density after nucleation, the
universe will be curvature dominated from the
beginning and never recover to a hot FRW universe.
Thus the entropy problem cannot be solved.
To solve this problem, a secondary inflation in the
bubble is required.
The essential difference of this scenario from the standard scenario
is that a horizon size patch whose size is approximately
equal to the curvature scale at the onset of the secondary inflation
 may not become much larger than the present horizon scale.
In such a case, we will have a sufficiently homogeneous open universe
at present. This implies that memories of the quantum state of
the universe at the beginning of the secondary inflation will not
be erased but will directly affect the observed large scale
temperature and density fluctuations\cite{bucher,Bruce,sataya2}.

Therefore the quantum state of a field, $\phi$,
inside the nucleated bubble, especially that of the inflaton field of
the secondary inflation, is to be examined.
A pioneering study of this subject was done by
Rubakov\cite{rubakov} and a formalism which respects the $O(4)$ symmetry
of the tunneling background was developed by Vachaspati and Vilenkin
\cite{vacha}. Meanwhile we developed a formalism based on the
multi-dimensional tunneling wave function\cite{tasaya} and applied it
to the $O(4)$-symmetric bubble nucleation without gravitational
effects\cite{styyPTP} and with gravitational effects\cite{tanaka}.
However, all of these previous works remained in a rather formal
level in the sense that techniques to investigate the quantum state
which have practical applicability to a general situation have not
been developed. Recently,
we have succeeded in giving a general and practical method to
obtain the quantum state inside a nucleated bubble in flat
space\cite{yamamoto} and its application to the case of the
quantum state of the tunneling field itself is in progress\cite{hamazaki}.
Thus an extension of this method to the case when gravity
comes into play is now to be formulated.

As a first step, we consider the case when the
gravitational back reaction effect can be neglected so that
the background metric may be fixed to that of de Sitter
space.
Then the spacetime inside a bubble is described by a spatially open
chart of the de Sitter space and the inflaton field will take a
constant value on a hypersurface of the spatially open time slicing.
Further, if the tunneling field which causes the bubble nucleation
has no interaction with $\phi$,
the quantum state of $\phi$ will not be
affected by the tunneling process at all.
Then, provided that the quantum state of $\phi$ before tunneling
is in the Euclidean vacuum, which should be a good approximation
if the preceding false vacuum inflation lasted long enough,
the quantum state inside a nucleated bubble will remain so.
Hence, it is required to describe the quantum
state by the mode functions in an open chart of de Sitter space.
Once we know a method to obtain the Euclidean vacuum mode functions
in the open chart, it should be fairly straightforward to extend it
to the case in which the mass of $\phi$ changes in time due to
its coupling to the tunneling field or when the geometry deviates
 from the exact de Sitter space at later stages.

In a spatially flat or closed chart, the field operator can be
relatively easily decomposed into the spatial harmonics and the mode
functions corresponding to the Euclidean vacuum are well
known\cite{BunDav}.
However, the Euclidean vacuum mode functions
in a spatially open chart have not been known except for the massless
conformally coupled case\cite{Pfa}.
Here we give an explicit expression of vacuum mode functions for
a scalar field with arbitrary mass and curvature coupling,
including the massless minimal coupling limit.

The paper is organized as follows. In Section II, we quantize
a scalar field on the hyperbolic time slices which foliate
two distinct open charts of de Sitter space (see Fig.~1).
We define a vacuum state there by requiring
that the positive frequency functions be regular on a hemisphere of
the Euclidean de Sitter space where the complexified time coordinate
is negative pure imaginary. Then we derive an expression for the
Wightman function for this vacuum state in the series form.
In Section III, by analyzing the behavior
of the Wightman function for thus obtained vacuum state,
we show that it is in fact the de Sitter-invariant
Euclidean vacuum, the so-called Bunch-Davies vacuum\cite{BunDav},
 provided that the effective mass of the scalar field is greater
than a critical value which corresponds to the conformally
coupled massless case. However, we also find that our expression for
the Wightman function does not coincide with the Euclidean vacuum one
for mass smaller than the critical value.
In Section IV, by carefully analyzing this discrepancy in the small
mass case, we find that there exist a set of modes
which have finite Klein-Gordon norms on regular
Cauchy surfaces, say on spatially closed time slices,
but which cannot be quantized on the open charts
because of the divergent Klein-Gordon norms on the
hyperbolic time slices. Then normalizing these modes on a spatially
closed hypersurface and analytically continuing them back to the
 open charts, we obtain a complete description of a scalar field
and the Euclidean vacuum there in terms of the orthonormalized
positive frequency functions, irrespective of its mass.
In Section V, we consider the special cases of a massless
conformal scalar and a massless minimal scalar
in which the series expression for the Wightman function can be
summed up with elementary algebra. The results are found to be in perfect
agreement with our analysis for general mass in the
preceding sections.
In Section VI, we summarize our results and discuss their
implications.
Finally, Appendix A explicitly evaluates the Klein-Gordon norms
of the mode functions on a closed slice for which
their norms on a hyperbolic slice are finite, to show their
equivalence and Appendix B gives a proof of a mathematical
formula which plays a central role in the analysis of the set
of modes whose Klein-Gordon norms diverge on a hyperbolic slice.

\section{Scalar field on open de Sitter space}
To begin with, we introduce a coordinate system which
covers the whole Euclidean de Sitter space.
The four-dimensional Euclidean de Sitter space is a four-sphere
and can be embedded in the five-dimensional Euclidean space,
$(\tilde x^0,x^i)$, as a hypersurface which satisfies
$H^{-2}=(\tilde x^0)^2+\sum (x^i)^2$,
where $H^{-1}$ is the Hubble radius of the de Sitter space.
A fundamental coordinate system we use in the Euclidean region is
defined as
\begin{equation}
 \tilde x^0 = \cos \tau \cos \rho,\quad x^1 = \sin \tau, \quad
 \left(\begin{array}{c}
  x^{2} \\
  x^{3} \\
  x^{4}
       \end{array}\right)
   =\cos\tau\sin\rho
 \left(\begin{array}{c}
    \cos\theta \\
    \sin\theta\cos\varphi \\
    \sin\theta\sin\varphi
    \end{array}\right).
\quad\left(\begin{array}{c}
    -\pi/2\le\tau\le\pi/2 \\
    0\le\rho\le\pi
    \end{array}
    \right)
\end{equation}
Then the metric is represented as
\begin{equation}
 ds^2_E=H^{-2}\left(
    d\tau^2+\cos^2 \tau(d\rho^2+\sin^2 \rho d\Omega^2)
    \right).
\end{equation}
Coordinate systems in the Lorentzian region are obtained by
the analytic continuation, $\tilde x^0\rightarrow ix^0$.
Performing this, we find that the Lorentzian region may be
divided into three parts, which we call $R$, $C$, and $L$,
whose coordinates are related to the fundamental ones by
the relations,
\begin{eqnarray}
&&  \left\{\begin{array}{rcll}
   t_{R} & = & i(\tau-\pi/2),\quad &(t_{R}\ge 0) \\
   r_{R} & = & i \rho,\quad &(r_{R}\ge 0)
  \end{array}\right.
 \nonumber \\
&&  \left\{\begin{array}{rcll}
   t_C & = & \tau,\quad &(\pi/2\ge t_{C}\ge -\pi/2) \\
   r_C & = & i(\rho-\pi/2),\quad &(\infty>r_C>-\infty)
  \end{array}\right.
 \nonumber \\
&&  \left\{\begin{array}{rcll}
   t_{L} & = & i(-\tau-\pi/2),\quad &(t_{L}\ge 0) \\
   r_{L} & = & i \rho,\quad &(r_{L}\ge 0)
  \end{array}\right.
 \label{connection}
\end{eqnarray}
and their metrics are given respectively by
\begin{eqnarray}
 && ds^2_{R} =H^{-2}\left(
    -dt^2_{R}+\sinh^2 t_{R}(dr_{R}^2+\sinh^2 r_{R}
     d\Omega^2)
    \right),
 \nonumber \\
 && ds^2_C =H^{-2}\left(
    dt^2_C+\cos^2 t_C (-dr_C^2+\cosh^2 r_C d\Omega^2)
    \right),
\\
 && ds^2_{L} =H^{-2}\left(
    -dt^2_{L}+\sinh^2 t_{L}(dr_{L}^2+\sinh^2 r_{L}
     d\Omega^2)
    \right).
 \nonumber \\
\end{eqnarray}
In the above, the regions $R$ and $L$ are connected through
$\tau\in(-\pi/2,\pi/2)$ along a path over the hemisphere
$\Im x^0<0$ ($\tilde x^0>0$), say along $\rho=0$, and
the region $C$ is joined to this hemisphere
at $\rho=\pi/2$. Note that one may connect the regions $R$ and $L$ in
a different way by passing through the other hemisphere $\Im x^0>0$
($\tilde x^0<0$), say along $\rho=\pi$, but we take the former
choice given by Eq.~(\ref{connection}) for later convenience.
A schematic picture of how these coordinates cover the spacetime is
shown in a conformal diagram in Fig.~1.
The regions $R$ and $L$ covered by the coordinates $(t_R,r_R)$ and
$(t_L,r_L)$, respectively, are the two distinct spatially open
charts of de Sitter space.
We see that the whole Euclidean and Lorentzian de Sitter space
is covered by complexifying the variables $\tau$ and $\rho$.

Let us consider a free scalar field in de Sitter space
with curvature coupling constant $\xi$.
We expand the field operator as
\begin{equation}
 \hat\phi(x)=\sum_\Lambda
 \left(\hat a_\Lambda u_\Lambda(x)
      +\hat a_\Lambda^{\dag} \overline{u_\Lambda(x)}
        \right),
\end{equation}
where a bar denotes the complex conjugate and $\{u_\Lambda(x)\}$ forms
a complete set of mode functions labeled by certain indices
$\Lambda$ which satisfy the field equation,
\begin{equation}
 \left[g^{\mu\nu}\nabla_{\mu}\nabla_{\nu}
   -M^2_{eff}\right] u_\Lambda(x)=0; \quad M^2_{eff}:=M^2+12\xi H^2,
 \label{feq}
\end{equation}
and normalized with respect to the Klein-Gordon inner product.
The condition, $\hat a_\Lambda\vert 0\rangle =0$ for any $\Lambda$,
determines the vacuum state associated with
a specified set of mode functions.

To find the Euclidean vacuum mode functions on the open chart,
we write down the field equation (\ref{feq}) in terms of the
coordinates in either of the regions $R$ or $L$.
Since these two regions are completely symmetric we have
\begin{equation}
\left[{1\over a^3(t)}{\partial\over\partial t}a^3(t)
{\partial\over\partial t}
-{H^{-2} \over a^2(t)} {\bf L}^2+{9\over 4}-\nu^2 \right]
u_\Lambda(t,r,\Omega)=0,
\label{apndxaa}
\end{equation}
where $(t,r)=(t_R,r_R)$ or $(t_L,r_L)$,
\begin{eqnarray}
a(t)&=&H^{-1}\sinh t,\quad
 \nu=\sqrt{{9\over 4}-{M^2_{eff}\over H^2}},
 \nonumber \\
{\bf L}^2&=&{1\over \sinh^2 r}{\partial\over\partial r}
\Biggl(\sinh^2 r{\partial\over\partial r}\Biggr)+
{1\over \sinh^2 r} {\bf L}^2_{\Omega},
\end{eqnarray}
and ${{\bf L}^2_{\Omega}}$ is the usual Laplacian operator on the
unit two-sphere.
We write the eigenvalue equation for the operator $-{\bf L}^2$ on a
unit three-dimensional hyperboloid in the form\cite{Gromes},
\begin{equation}
-{\bf L}^2 Y_{plm}(r,\Omega)=(1+p^2)Y_{plm}(r,\Omega).
\end{equation}
The eigenfunction $Y_{plm}$ which is regular at $r=0$ is given by
\begin{eqnarray}
Y_{plm}(r,\Omega)&=&f_{pl}(r)Y_{lm}(\Omega),
\nonumber\\
 f_{pl}(r)&:=&
{\Gamma(ip+l+1)\over\Gamma(ip+1)}{p\over\sqrt{\sinh r}}
   P^{-l-1/2}_{ip-1/2}(\cosh r)
\nonumber\\
&=&
(-1)^l\sqrt{2\over\pi}\,{\Gamma(-ip+1)\over\Gamma(-ip+l+1)}\,
  \sinh^lr{d^l\over d(\cosh r)^l}
    \left({\sin pr\over\sinh r}\right).
\label{fpl}
\end{eqnarray}
where $Y_{lm}(\Omega)$ is the normalized
spherical harmonic function on the unit two-sphere,
$\Gamma(z)$ is the Gamma function and $P^{\nu}_{\mu}(z)$ is
the associated Legendre function of the first kind\cite{Magnus}.
The eigenfunctions $Y_{plm}$ with real positive values of $p$
form an orthonormal complete set for square-integrable
functions on the unit three-hyperboloid.
They are normalized as
\begin{equation}
\int_0^\infty dr\sinh^2r\int d\Omega\, Y_{plm}(r,\Omega)
\overline{Y_{p'l'm'}(r,\Omega)}
  =\delta(p-p')\delta_{ll'}\delta_{mm'}\,.
\label{NormYplm}
\end{equation}
We now take the harmonic expansion of the mode functions,
\begin{equation}
u_{plm}(t,r,\Omega)=
{1\over a(t)}\chi_{plm}(t) Y_{plm}(r,\Omega),
 \label{sep}
\end{equation}
and look for a complete set of positive frequency functions
$\{\chi_{plm}(t)\}$ for the Euclidean vacuum.

As well-known, the positive frequency functions
of the Euclidean vacuum are characterized by their regularity on
the $\Im x^0<0$ hemisphere of the Euclidean de Sitter space.
To find such functions, let us first consider the mode functions in
the region $R$.
The general solution is expressed as a linear combination
of the associated Legendre functions $P^{ip}_{\nu-1/2}(z_R)$ and
$P^{-ip}_{\nu-1/2}(z_R)$, where $z_R=\cosh t_R$.
To find out their coefficients, we first ignore the normalization and
consider a solution,
\begin{equation}
  \chi_p^{(R)}=  P^{ip}_{\nu'}(z_R).
\end{equation}
where we have introduced $\nu':=\nu-1/2$ for notational simplicity.
Note that this function would be a natural candidate for the positive
frequency function if the region $R$ were the whole universe.
Imposing the above regularity condition on $\chi_p^{(R)}$,
we find that the analytic continuation of it to
the region $L$ amounts to going through the branch cut $[-1,1]$
of the associated Legendre function $P^{ip}_{\nu'}(z)$ from
$\Im z<0$ to $\Im z>0$ and then to $z<-1$ on the real axis.
Thus in the region $L$ we have
\begin{eqnarray}
  \chi_p^{(R)} & = &
  e^{-\pi p} P^{ip}_{\nu'}(-z_L)
  \nonumber \\
  & = &
  e^{\pi (-p+i\nu')} P^{ip}_{\nu'}(z_L)
   - {2\sin\pi(ip+\nu')\over\pi}Q^{ip}_{\nu'}(z_L),
\end{eqnarray}
where $z_L=\cosh t_L$ and $Q^{~\mu}_{\nu}(z)$ is
the associated Legendre function of the second kind\cite{Magnus}.
Replacing the role of $R$ and $L$, we obtain $\chi_p^{(L)}$
in the same way. With this choice of the mode functions, their
regularity on the $\Im x^0<0$ hemisphere is automatically guaranteed
because the harmonic function $Y_{plm}$ is regular at
$r_R=r_L=\rho=0$, hence regular for $0\le\rho\le\pi/2$.

As the mode functions we have obtained are not yet normalized,
the Klein-Gordon inner products of these mode functions should be
calculated to normalize them.
The evaluation of the Klein-Gordon
inner products must be performed on a Cauchy surface.
As the choice of a surface is arbitrary as long as it is a Cauchy
surface, we choose it in the following way.
It consists of the three parts, (I),(II) and (III), where
(I) is the $r_R<r_{max}$ part of $t_R={\rm constant}$ hypersurface
for a large $r_{max}$,
(II) is the one in which $R$ is replaced by $L$, and
(III) is a bridge connecting these two isolated parts.
A schematic picture of this Cauchy surface is drawn in Fig.~2.
If the contribution from the integral over the surface (III)
can be neglected for sufficiently large $r_{max}$,
the Klein-Gordon inner product reduces to the summation of integrals
over the surfaces (I) and (II). In such a case we have
\begin{eqnarray}
 \Bigl\langle{\chi_1(t) \over a(t)} Y_{plm}(r,\Omega)\,
&&,\,{\chi_2(t) \over a(t)}Y_{p'l'm'}(r,\Omega)\Bigr\rangle
\nonumber\\
&&=\left[\left\{i(z^2_R-1)\left({d\chi_1 \over dz_R}\overline{\chi_2}
 - \chi_1 \overline{d\chi_2  \over dz_R}\right)\right\}
+\{R\rightarrow L\} \right]
 \delta(p-p')\delta_{ll'} \delta_{mm'}
\nonumber\\
&&=:(\!(\chi_1,\chi_2)\!)\,\delta(p-p')\delta_{ll'} \delta_{mm'}\,.
 \label{KGip}
\end{eqnarray}
As will be discussed in Section IV, this is not always correct
but we assume so for the time being.

Then the Klein-Gordon inner products of $\chi_p^{(R)}$ and
$\chi_p^{(L)}$ are calculated to be
\begin{eqnarray}
&& (\!(\chi_p^{(R)},\chi_p^{(L)})\!)
  =(\!(\chi_p^{(L)},\chi_p^{(R)})\!)=0,
\nonumber \\
&& (\!(\chi_p^{(R)},\chi_p^{(R)})\!)
  =(\!(\chi_p^{(L)},\chi_p^{(L)})\!)
  = {2\over\pi}e^{-\pi p}(\cosh 2\pi p -\cos 2\pi\nu'),
\label{KGnorm}
\end{eqnarray}
where we have used the fact that
\begin{equation}
\overline{P^{ip}_{\nu'}}=P^{-ip}_{\nu'}\,,
\qquad
\overline{Q^{ip}_{\nu'}}=e^{-2\pi p}Q^{-ip}_{\nu'}\,,
\label{CCofPQ}
\end{equation}
and the Wronskian relations among $P^{\pm ip}_{\nu'}$ and
$Q^{\pm ip}_{\nu'}$ \cite{Magnus}. In the above and in the rest of
this section, we assume $\nu'$ to be real,
i.e., $-1/2\le\nu'<1$ ($0\le\nu<3/2$), in order to avoid inessential
complexity. Extension to the case $\nu'$ is imaginary, i.e.,
$\Re\nu'=-1/2$ ($\nu={\rm pure~imaginary}$) is straightforward.
It then turns out that the following linear combinations of
$\chi_p^{(R)}$ and $\chi_p^{(L)}$ are also mutually orthogonal:
\begin{eqnarray}
  \chi_{p,+}&=&
 {\chi_p^{(R)}+\chi_p^{(L)}
  \over(1+e^{-\pi p+\nu'\pi i})\Gamma(\nu'+1+ip)}
\nonumber\\
&=&\left\{\begin{array}{l}
\displaystyle {1\over\Gamma(\nu'+1+ip)}
 \left[P^{ip}_{\nu'}(z_R)
       +{i\over\pi}(1-e^{p\pi-i\nu'\pi})Q^{ip}_{\nu'}(z_R)\right],
\\\\
\displaystyle {1\over\Gamma(\nu'+1+ip)}
 \left[P^{ip}_{\nu'}(z_L)
       +{i\over\pi}(1-e^{p\pi-i\nu'\pi})Q^{ip}_{\nu'}(z_L)\right],
 \end{array}\right.
\nonumber\\\nonumber\\
  \chi_{p,-}&=&
 {\chi_p^{(R)}-\chi_p^{(L)}
  \over(1-e^{-\pi p+\nu'\pi i})\Gamma(\nu'+1+ip)}
\nonumber\\
&=&\left\{\begin{array}{l}
\displaystyle{1\over\Gamma(\nu'+1+ip)}
 \left[P^{ip}_{\nu'}(z_R)
       +{i\over\pi}(1+e^{p\pi-i\nu'\pi})Q^{ip}_{\nu'}(z_R)\right],
\\\\
\displaystyle -{1\over\Gamma(\nu'+1+ip)}
 \left[P^{ip}_{\nu'}(z_L)
       +{i\over\pi}(1+e^{p\pi-i\nu'\pi})Q^{ip}_{\nu'}(z_L)\right].
 \end{array}\right.
\label{modefcn}
\end{eqnarray}
If we note the relation,
\begin{equation}
Q^{ip}_{\nu'}={\pi e^{-\pi p}\over2i\sinh\pi p}\left[P^{ip}_{\nu'}
-{\Gamma(\nu'+1+ip)\over\Gamma(\nu'+1-ip)}P^{-ip}_{\nu'}\right],
\label{QtoP}
\end{equation}
we may re-express $\chi_{p,\sigma}$ in terms of $P^{ip}_{\nu'}$
alone as
\begin{equation}
 \chi_{p,\sigma} =
 \left\{
 \begin{array}{l}
  \displaystyle
 {1\over 2\sinh\pi p}\left(
 {e^{\pi p}-\sigma e^{-i\pi\nu'}\over \Gamma(\nu'+ip +1)}
 P^{ip}_{\nu'}(z_R)-
 {e^{-\pi p}-\sigma e^{-i\pi\nu'}\over \Gamma(\nu'-ip +1)}
 P^{-ip}_{\nu'}(z_R)\right),
 \\\\
  \displaystyle
 {\sigma\over 2\sinh\pi p}\left(
 {e^{\pi p}-\sigma e^{-i\pi\nu'}\over \Gamma(\nu'+ip +1)}
 P^{ip}_{\nu'}(z_L)-
 {e^{-\pi p}-\sigma e^{-i\pi\nu'}\over \Gamma(\nu'-ip +1)}
 P^{-ip}_{\nu'}(z_L)\right).
 \end{array}
 \right.
\label{modefcn2}
\end{equation}
Note that these mode functions have the symmetry
$\chi_{-p,\sigma}=\chi_{p,\sigma}$ ($\sigma=\pm$), hence are even
functions of $p$.
Their Klein-Gordon norms are evaluated to give
\begin{equation}
 (\!(\chi_{p,\sigma},{\chi_{p,\sigma'}})\!)
  ={4\left(\cosh\pi p-\sigma\cos\nu'\pi\right)\over\pi
       \left|\Gamma(\nu'+1+ip)\right|^2} \delta_{\sigma\sigma'}
  =: N_{p\sigma}\delta_{\sigma\sigma'}\,.
\label{chinorm}
\end{equation}
{}From these results, the field operator is now expanded as
\begin{equation}
\hat\phi(x)=\int_0^\infty dp\sum_{\sigma,l,m}
 \left(\hat a_{p\sigma lm}v_{p\sigma lm}(x)+
     \hat a_{p\sigma lm}^{\dag}\overline{v_{p\sigma lm}(x)}\right),
\end{equation}
where and in what follows, we use the symbol $v_\Lambda$ to denote
the orthonormalized mode functions. In the present case, they are
given by
\begin{equation}
 v_{p\sigma lm}(x)={1\over\sqrt{N_{p\sigma}}}
           {\chi_{p,\sigma}(t)\over a(t)}f_{pl}(r)Y_{lm}(\Omega),
\label{Normmode}
\end{equation}
with $(t,r)$ being either $(t_R,r_R)$ or $(t_L,r_L)$.
In what follows, we suppress the subscript $R$ (or $L$)
for notational simplicity unless ambiguity arises.

Then, for both $x'$ and $x''$ in the region $R$ (or $L$),
the Wightman function is given by
\begin{eqnarray}
 a(t')a(t'')G^{+}(x',x'')
&=&
 \int_{0}^{\infty}dp
    \sum_{\sigma,l,m}{1\over N_{p\sigma}}\,
  \chi_{p,\sigma}(t')Y_{plm}(r',\Omega')
 \overline{\chi_{p,\sigma}(t'')Y_{plm}(r'',\Omega'')}\,,
\nonumber\\
&=&
 {1\over 2\pi^2}
 \int_{0}^{\infty}dp\,{p\,\sin p\zeta\over\sinh\zeta}
    \sum_{\sigma=\pm}{1\over N_{p\sigma}}\,
  \chi_{p,\sigma}(t')\overline{\chi_{p,\sigma}(t'')}\,,
 \label{oriG}
\end{eqnarray}
where we have used the completeness relation for $Y_{plm}$,
\begin{eqnarray}
\sum_{l,m}&&Y_{plm}(r',\Omega')\overline{Y_{plm}(r'',\Omega'')}
 ={p\,\sin p\zeta\over2\pi^2\sinh\zeta}\,;
\nonumber\\
&&\cosh\zeta:=\cosh r'\cosh r''-\sinh r'\sinh r''\cos\Theta,
\label{completeness}
\end{eqnarray}
with $\cos\Theta$ being the directional cosine
 between $\Omega'$ and $\Omega''$.
Using the relations (\ref{CCofPQ}) and (\ref{QtoP}),
and the symmetry $\chi_{-p,\sigma}=\chi_{p,\sigma}$ again,
the above integral is written explicitly as
\begin{eqnarray}
 a(t')a(t'')G^{+}(x',x'')
&=&{1\over 8\pi\sinh\zeta}\int_{-\infty}^{\infty}dp\,
 {p\sin p\zeta\over\sinh\pi p}
\nonumber\\
\nonumber\\
&&\quad\times\Biggl[{ie^{\pi(p-i\nu')}\over\sin(ip+\nu')\pi}
 P^{ip}_{\nu'}(z')P^{-ip}_{\nu'}(z'')
-{2i\sin\nu'\pi\over\pi\sin(ip+\nu')\pi}e^{-\pi p}
 P^{ip}_{\nu'}(z')Q^{-ip}_{\nu'}(z'')\Biggr]
\nonumber\\
\nonumber\\
&=&{1\over 8\pi\sinh\zeta}{1\over2\pi i}\int_{-\infty}^{\infty}
 idp\,{\pi p\over\sinh\pi p}\,e^{ip\zeta}
\nonumber\\
\nonumber\\
&&\quad\times\Biggl[{e^{\pi(p-i\nu')}\over\sin(ip+\nu')\pi}
 P^{ip}_{\nu'}(z')P^{-ip}_{\nu'}(z'')
-{2\sin\nu'\pi\over\pi\sin(ip+\nu')\pi}e^{-\pi p}
 P^{ip}_{\nu'}(z')Q^{-ip}_{\nu'}(z'')
\nonumber\\
\nonumber\\
&&\qquad+{e^{-\pi(p+i\nu')}\over\sin(ip-\nu')\pi}
 P^{-ip}_{\nu'}(z')P^{ip}_{\nu'}(z'')
-{2\sin\nu'\pi\over\pi\sin(ip-\nu')\pi}e^{\pi p}
 P^{-ip}_{\nu'}(z')Q^{ip}_{\nu'}(z'')\Biggr].
\label{Wightman}
\end{eqnarray}
In deriving the above expression, we have assumed $\nu=\nu'+1/2$
to be real. However, it can be shown that the final result (\ref{Wightman})
equally holds for imaginary $\nu$ without change. Note that
this is true irrespective of the sign of $\nu$,
i.e., for $\nu=\pm i|\nu|$.
Note also that $p=0$ is no longer a pole for $\nu'\neq$integer,
which we assume in the rest of this section.
The cases $\nu'=0$ ($\nu=1/2$; massless conformal) and
$\nu'=1$ ($\nu=3/2$; massless minimal) will be
treated separately in Section V.

When $x'$ and $x''$ are spatially separated,
we may close the contour of integration over the upper half complex
$p$-plain. Changing the integration variable to $u=ip$,
Eq.~(\ref{Wightman}) may be rewritten in the form,
\begin{eqnarray}
 a(t')a(t'')G^{+}(x',x'')
 & = &{1\over 8\pi \sinh \zeta}{1\over 2\pi i}
   \int_C du\,{\pi u\over \sin\pi u}\, e^{-u\zeta}
 \nonumber \\
 && \left\{{e^{i\pi(u-\nu')}\over\sin\pi(u-\nu')}\left(
   P^{-u}_{\nu'}(z') P^{u}_{\nu'}(z'')
   -{2\over\pi}\sin\pi\nu'e^{i\pi(\nu'-2u)}
   P^{-u}_{\nu'}(z') Q^{u}_{\nu'}(z'')\right)\right.
 \nonumber \\
 && \quad +
   \left.{e^{-i\pi(u+\nu')}\over\sin\pi(u+\nu')}\left(
   P^{u}_{\nu'}(z') P^{-u}_{\nu'}(z'')
   -{2\over\pi}\sin\pi\nu'e^{i\pi(\nu'+2u)}
   P^{u}_{\nu'}(z') Q^{-u}_{\nu'}(z'')\right)\right\},
\label{contourint}
\end{eqnarray}
where the contour of integration $C$ on the complex $u$-plane
is shown in Fig.~3.
If we note the relations,
\begin{equation}
 P^{-n}_{\nu'}(z') P^{n}_{\nu'}(z'')
 =P^{n}_{\nu'}(z') P^{-n}_{\nu'}(z''), \quad
 P^{-n}_{\nu'}(z') Q^{n}_{\nu'}(z'')
 =P^{n}_{\nu'}(z') Q^{-n}_{\nu'}(z''),
\end{equation}
for integer $n$,
we easily see that the residues of poles at $u=n$ ($n=1,2,3,\cdots$)
cancel out completely.
Thus the poles which contribute to the integral
are classified into three classes.
\medskip

\noindent
{\it class}$(a)$; poles of $1/\sin\pi(u-\nu')$:
\begin{equation}
    u=\nu'+n, \quad \left\{\begin{array}{l}
    n=1,2,3,\cdots,~ (-{1\over 2}\le\nu'<0, \Re\nu'=-1/2) \\
    n=0,1,2,\cdots.~ (0<\nu'<1)
    \end{array}\right.
\end{equation}

\noindent
{\it class}$(b)$; poles of $Q^{-u}_{\nu'}(z'')$:
\begin{equation}
    u=\nu'+n,  \quad
    ~n=1,2,3,\cdots.
\end{equation}

\noindent
{\it class}$(c)$; poles of $1/\sin\pi(u+\nu')$:
\begin{equation}
    u=-\nu'+n, \quad \left\{\begin{array}{l}
    n=0,1,2,\cdots,~ (-{1\over 2}\le\nu'<0, \Re\nu'=-1/2) \\
    n=1,2,3,\cdots.~ (0<\nu'<1)
    \end{array}\right.
\end{equation}

The contribution from a {\it class} $(a)$ pole is given by
\begin{equation}
 a_n:={1\over 8\pi\sinh\zeta}
 {(-1)^n\over\sin\pi\nu'} (\nu'+n) e^{-(\nu'+n)\zeta}
  \left(
  P^{-(\nu'+n)}_{\nu'}(z') P^{\nu'+n}_{\nu'}(z'')
  -{2\over\pi}\sin\pi\nu'e^{-i\pi\nu'}
  P^{-(\nu'+n)}_{\nu'}(z') Q^{\nu'+n}_{\nu'}(z'')\right) .
\end{equation}
By using the following formulae,
\begin{eqnarray}
 Q^{\nu'+n}_{\nu'}(z) &= &
 {\pi\over 2}{e^{i\pi\nu'}\over\sin\pi\nu'}P^{\nu'+n}_{\nu'}(z);
 \quad n=1,2,3,\cdots,
\nonumber \\
 Q^{\nu'-n}_{\nu'}(z) & =&
 {\pi\over 2}{e^{i\pi\nu'}\over\sin\pi\nu'}\left(P^{\nu'-n}_{\nu'}(z)
 -{\Gamma(2\nu'-n+1)\over n!}P^{-\nu'+n}_{\nu'}(z)\right);
 \quad n=0,1,2,\cdots,
 \label{QtoP2}
\end{eqnarray}
we find
\begin{eqnarray}
a_0 & = & {1\over 8\pi\sinh\zeta}
   {\nu'\Gamma(2\nu'+1)\over\sin\pi\nu'}
   e^{-\nu'\zeta}P^{-\nu'}_{\nu'}(z') P^{-\nu'}_{\nu'}(z''),
\nonumber \\
a_n & = & 0;\quad n=1,2,3,\cdots.
\end{eqnarray}

The contribution from a {\it class} $(b)$ pole is given by
\begin{equation}
 b_n:=-{1\over 8\pi\sinh\zeta}
 {(-1)^n \pi\over\sin\pi\nu'} (\nu'+n) e^{-(\nu'+n)\zeta}
 {2\over\pi}{e^{i\pi\nu'}\sin\pi\nu'\over\sin 2\pi\nu'}
 P^{\nu'+n}_{\nu'}(z') \left[\lim_{u\rightarrow\nu'+n}
 (u-\nu'-n)Q^{-u}_{\nu'}(z'')\right].
\end{equation}
The limit in the square bracket is evaluated as
\begin{eqnarray}
 \lim_{u\rightarrow\nu'+n} (u-\nu'-n)Q^{-u}_{\nu'}(z'')
 && =\lim_{u\rightarrow\nu'+n}
  (u-(\nu'+n))\Gamma(\nu'-u+1){Q^{-u}_{\nu'}(z'')\over
 \Gamma(\nu'-u+1)}
 \nonumber \\
 && ={(-1)^n\over (n-1)!}{\pi\over 2}{e^{-i\pi\nu'}\over
 \sin\pi\nu' \Gamma(2\nu'+n+1)} P^{\nu'+n}_{\nu'}(z'').
\end{eqnarray}
Then we obtain
\begin{equation}
 b_n=-{1\over 8\pi\sinh\zeta}
   {\pi (\nu'+n) e^{-(\nu'+n)\zeta}
     \over (n-1)!\sin\pi\nu' \sin 2\pi\nu' \Gamma(2\nu'+n+1)}
     P^{\nu'+n}_{\nu'}(z')
        P^{\nu'+n}_{\nu'}(z'');\quad n=1,2,3,\cdots.
\end{equation}

The contribution from a {\it class} $(c)$ pole is given by
\begin{equation}
 c_n:={1\over 8\pi\sinh\zeta}
 {(-1)^n\over\sin\pi\nu'} (\nu'-n) e^{(\nu'-n)\zeta}
 \left(P^{-\nu'+n}_{\nu'}(z') P^{\nu'-n}_{\nu'}(z'')
  -{2\over\pi}\sin\pi\nu'e^{-i\pi\nu'}
  P^{-\nu'+n}_{\nu'}(z') Q^{\nu'-n}_{\nu'}(z'')\right).
\end{equation}
Using the second formula in Eq.~(\ref{QtoP2}), it reduces to
\begin{equation}
 c_n={1\over 8\pi\sinh\zeta}
  {(-1)^n\over \sin\pi\nu'}(\nu'-n) e^{(\nu'-n)\zeta}
  {\Gamma(2\nu'-n+1)\over n!}
  P^{-\nu'+n}_{\nu'}(z') P^{-\nu'+n}_{\nu'}(z'');\quad n=0,1,2,\cdots.
\end{equation}

Combining these results, we obtain the Wightman function expressed in
terms of the series sum of the residues.
Here we focus on the case when $-1/2\le\nu'<0$ or $\Re\nu'=-1/2$.
Then we have
\begin{eqnarray}
 G^{+}(x',x'')
 &=&  {1\over 8\pi a(t')a(t'')\sinh\zeta\sin\pi\nu'}
\left\{-{\pi\over\sin 2\pi\nu'}
   \sum_{n=1}^{\infty}(\nu'+n) A_n
   +\sum_{n=0}^{\infty}(\nu'-n) B_n\right\}
\nonumber\\
 &=:&G^{+}_{(A)}(x',x'')+G^{+}_{(B)}(x',x''),
\label{Green}
\end{eqnarray}
where
\begin{eqnarray}
 A_n &:=&{e^{-(\nu'+n)\zeta} \over (n-1)! \Gamma(2\nu'+n+1)}
     P^{\nu'+n}_{\nu'}(z') P^{\nu'+n}_{\nu'}(z''),
 \nonumber \\
 B_n &:=&{(-1)^n e^{(\nu'-n)\zeta} \Gamma(2\nu'-n+1) \over n!}
     P^{-\nu'+n}_{\nu'}(z') P^{-\nu'+n}_{\nu'}(z''),
\end{eqnarray}
and $G^{+}_{(A)}(x',x'')$ and $G^{+}_{(B)}(x',x'')$ correspond to the
summations of $A_n$ and $B_n$ parts, respectively.

In the next section, we shall show that the above expression
indeed coincides with the Euclidean vacuum Wightman function.
The case $\nu'>0$ will be discussed in Section IV.

\section{Euclidean vacuum Wightman function}

 Equation~(\ref{Green}) has been derived by assuming that the
Klein-Gordon inner product can be evaluated in terms of the expression
given by Eq.~(\ref{KGnorm}). If this assumption is valid,
we expect Eq.~(\ref{Green}) to coincide with the
well-known Wightman function for the Euclidean vacuum\cite{BunDav},
\begin{equation}
 G^{+}_E ({z',0;z'',\zeta})
   ={H^2\over 8\pi^2}\Gamma\left({3\over2}+\nu\right)
    \Gamma\left({3\over2}-\nu\right)
    {P^{-1}_{\nu'}(u)\over \sqrt{u^2-1}},
  \label{gWeit}
\end{equation}
where
\begin{equation}
 u:=s' s'' \cosh\zeta -z'z'' ;\quad s=\sqrt{z^2-1}=\sinh t.
\end{equation}

Now we prove that Eq.~(\ref{Green})
is identical to the Euclidean vacuum Wightman function.
For convenience, we denote the function given by the series
expression (\ref{Green}) by $\tilde G^{+}(x',x'')$.
The outline of the proof is as follows.
First we show that $\tilde G^{+}(x',x'')$ is de Sitter-invariant,
i.e., it is a function of the invariant quantity $u$ alone.
Then, since $\tilde G^{+}(x',x'')$ is guaranteed to be a solution
of the field equation
with respect to both of the arguments $x'$ and $x''$ by construction,
the form of $\tilde G^{+}(z',z'',\zeta)$ is restricted to be a
linear combination of two independent solutions as
\begin{equation}
 \tilde G^{+}(z',z'',\zeta)=
 {H^2 \over 8\pi^2}{\Gamma(\nu'+2) \Gamma(-\nu'+1) \over
 \sqrt{u^2-1}}\left(\alpha P^{-1}_{\nu'}(u)
 +\beta{\tan\pi\nu'\over\pi}Q^{-1}_{\nu'}(u)\right),
\label{deSinv}
\end{equation}
where $\alpha$ and $\beta$ are some constants.
Examining the asymptotic behaviour of the Wightman function
for the limit of large spatial separation, $\alpha$ and
$\beta$ are determined as $\{\alpha =1,\beta=0\}$, which
is of the Euclidean vacuum Wightman function.

To show the de Sitter invariance of $\tilde G^{+}(z',z'',\zeta)$,
we use the fact that
if a function $F(z',z'',\zeta)$ satisfies the relations,
\begin{eqnarray}
 {\partial u\over \partial\zeta}
  {\partial F(z',z'',\zeta)\over \partial z'}
& = &
 {\partial u\over \partial z'}
  {\partial F(z',z'',\zeta)\over \partial\zeta},
\label{condition1}
 \\
 {\partial u\over \partial\zeta}
  {\partial F(z',z'',\zeta)\over \partial z''}
& = &
 {\partial u\over \partial z''}
  {\partial F(z',z'',\zeta)\over \partial\zeta},
 \label{condition2}
\end{eqnarray}
$F(z',z'',\zeta)$ is a function of $u$ only, i.e.,
$F(z',z'',\zeta)$ is a de Sitter-invariant function.
Of course we can directly check the relations (\ref{condition1})
and (\ref{condition2}) for $F(z',z'',\zeta)=\tilde G^{+}(z',z'',\zeta)$,
 but it is easier to show these relations first for
\begin{equation}
 F(z',z'',\zeta)={1\over\sin\pi\nu'}\left({\pi\over\sin 2\pi\nu'}
   \sum_{n=1}^{\infty} A_n+ \sum_{n=0}^{\infty} B_n\right).
\end{equation}
Then if $F(z',z'',\zeta)$ is shown to be a function of $u$ alone,
$\tilde G^{+}(z',z'',\zeta)$ can be obtained as
\begin{equation}
 8 \pi \tilde G^{+}(z',z'',\zeta)
 ={1\over s's''\sinh\zeta}{\partial\over \partial\zeta}F(u(z',z'',\zeta))
 ={d\over du} F(u).
\end{equation}
Hence $\tilde G^{+}(z',z'',\zeta)$ is also a function of only $u$ if
$F(z',z'',\zeta)$ is so.

Now let us prove the relations (\ref{condition1}) and
(\ref{condition2}). We define
\begin{eqnarray}
 K_n^{(1)} & := & {\partial u\over \partial z'}
{\partial A_n\over \partial\zeta}=
 -\left({s''z'\over s'} \cosh\zeta- z''\right)
   (\nu'+n)P^{\nu'+n}_{\nu'}(z')
   {e^{-(\nu'+n)\zeta} \over (n-1)! \Gamma(2\nu'+n+1)}
  P^{\nu'+n}_{\nu'}(z''),
 \nonumber \\
 K_n^{(2)} & := & {\partial u\over \partial\zeta}
{\partial A_n\over \partial z'}=
  {s''\over s'}\sinh\zeta\left((\nu'+n)z'P^{\nu'+n}_{\nu'}(z')
  +s'P^{\nu'+n+1}_{\nu'}(z')\right)
   {e^{-(\nu'+n)\zeta} \over (n-1)! \Gamma(2\nu'+n+1)}
  P^{\nu'+n}_{\nu'}(z''),
 \nonumber \\
 L_n^{(1)} & := & {\partial u\over \partial z'}
{\partial B_n\over \partial\zeta}=
  \left({s''z'\over s'} \cosh\zeta- z''\right)
   (\nu'-n) P^{-\nu'+n}_{\nu'}(z')
   {(-1)^n e^{(\nu'-n)\zeta}\Gamma(2\nu'-n+1) \over n!}
  P^{-\nu'+n}_{\nu'}(z''),
 \nonumber \\
 L_n^{(2)} & := & {\partial u\over \partial\zeta}
{\partial B_n\over \partial z'}=
  {s''\over s'}\sinh\zeta\left(-(\nu'-n)z'P^{-\nu'+n}_{\nu'}(z')
  +s'P^{-\nu'+n+1}_{\nu'}(z')\right)
\nonumber\\
&&\qquad\qquad\qquad\qquad\qquad\qquad
 \times {(-1)^n e^{(\nu'-n)\zeta} \Gamma(2\nu'-n+1) \over n!}
   P^{-\nu'+n}_{\nu'}(z'').
\end{eqnarray}
Then
\begin{eqnarray}
 2\sum_{n=1}^{\infty} \left( K_n^{(1)} - K_n^{(2)} \right)
  =
 \sum_{n=1}^{\infty}&&
 \Biggl[\left(-2{s''z'\over s'} e^{\zeta}+2z''\right)
 (\nu'+n)P^{\nu'+n}_{\nu'}(z')
  -s''(e^{\zeta}-e^{-\zeta})P^{\nu'+n+1}_{\nu'}(z')\Biggr]
  \nonumber \\  &&\quad
  \times  {e^{-(\nu'+n)\zeta}  \over
  (n-1)! \Gamma(2\nu'+n+1)} P^{\nu'+n}_{\nu'}(z'')
  \nonumber \\
   =  \sum_{n=0}^{\infty} && \Biggl[
  -{1\over n! \Gamma(2\nu'+n+2)}
  \left(2 (\nu'+n+1) {z'\over s'} P^{\nu'+n+1}_{\nu'}(z')
  +P^{\nu'+n+2}_{\nu'}(z')\right)
  s'' P^{\nu'+n+1}_{\nu'}(z'')
  \nonumber \\  &&\quad
  +2{\nu'+n \over (n-1)! \Gamma(2\nu'+n+1)}
  P^{\nu'+n}_{\nu'}(z') z''P^{\nu'+n}_{\nu'}(z'')
  \nonumber \\  &&\quad
  +{1\over (n-2)! \Gamma(2\nu'+n)}
  P^{\nu'+n}_{\nu'}(z') s''P^{\nu'+n-1}_{\nu'}(z'')\Biggr]
  e^{-(\nu'+n)\zeta}.
\end{eqnarray}
Each $n$ term in the above expression is shown to be zero by
iteratively using the recursion formula,
\begin{equation}
 P^{\nu'+n+2}_{\nu'}(z)
 +2(\nu'+n+1) {z\over s} P^{\nu'+n+1}_{\nu'}(z)
 +n(2\nu'+n+1) P^{\nu'+n}_{\nu'}(z)=0.
\end{equation}
In the same way, we have
\begin{eqnarray}
 2\sum_{n=0}^{\infty} \left( L_n^{(1)} - L_n^{(2)} \right)
  =  \sum_{n=0}^{\infty}
 && \Biggl[\left(2{s''z'\over s'} e^{\zeta}-2z''\right)
  (\nu'-n) P^{-\nu'+n}_{\nu'}(z')
  -s''(e^{\zeta}-e^{-\zeta})P^{-\nu'+n+1}_{\nu'}(z')\Biggr]
  \nonumber \\  &&\quad
  \times {(-1)^n e^{(\nu'-n)\zeta} \Gamma(2\nu'-n+1) \over
  n!} P^{-\nu'+n}_{\nu'}(z'')
  \nonumber \\
   =
  \sum_{n=-1}^{\infty} && \Biggl[
  -{\Gamma(2\nu'-n) \over (n+1)!}
  \left(2 (\nu'-n-1) {z'\over s'} P^{-\nu'+n+1}_{\nu'}(z')
  -P^{-\nu'+n+2}_{\nu'}(z')\right)
  s'' P^{\nu'+n+1}_{\nu'}(z'')
  \nonumber \\
  & & \quad  -2{(\nu'-n) \Gamma(2\nu'-n+1) \over n!}
  P^{-\nu'+n}_{\nu'}(z') z''P^{-\nu'+n}_{\nu'}(z'')
  \nonumber \\  &&\quad
  -{\Gamma(2\nu'-n+2) \over (n-1)! }
  P^{-\nu'+n}_{\nu'}(z') s''P^{-\nu'+n-1}_{\nu'}(z'')\Biggr]
  (-1)^n e^{(\nu'-n)\zeta},
\end{eqnarray}
and each $n$ term is shown to be zero by using the formula,
\begin{equation}
 P^{-\nu'+n+2}_{\nu'}(z)
 +2(-\nu'+n+1) {z\over s} P^{-\nu'+n+1}_{\nu'}(z)
 +(n+1)(-2\nu'+n) P^{-\nu'+n}_{\nu'}(z)=0.
\end{equation}
Thus we find that $F(z',z'',\zeta)$ satisfies the condition
(\ref{condition1}).
The condition (\ref{condition2}) is also shown to be
satisfied by replacing the roles of $z'$ and $z''$.
Hence we find that $\tilde G^{+}(z',z'',\zeta)$
is de Sitter-invariant as well as $F(z',z'',\zeta)$.

As mentioned before, since $\tilde G^{+}(z',z'',\zeta)$
is de Sitter-invariant, the fact that this function satisfies
the field equation restricts its form to that given by Eq.~(\ref{deSinv}).
Moreover, as is apparent from the above proof of the de Sitter invariance,
$G^{+}_{(A)}(z',z'',\zeta)$ and $G^{+}_{(B)}(z',z'',\zeta)$
are both de Sitter-invariant and satisfy the field equation
independently.
This means that these two functions give two different
de Sitter-invariant functions.
To determine $\alpha$ and $\beta$ for each of them,
we examine their asymptotic behaviours at large $\zeta$ limit.
In this limit, $2u\rightarrow s's''e^{\zeta}$.

The asymptotic behaviours of $G^{+}_{(A)}(z',z'',\zeta)$
and $G^{+}_{(B)}(z',z'',\zeta)$ are determined by their leading terms.
Noting that
\begin{equation}
 P^{\nu'+1}_{\nu'}(z) \rightarrow{2^{\nu'+1}\over \Gamma(-\nu')}
   s^{-\nu'-1}, \quad
 P^{-\nu'}_{\nu'}(z) \rightarrow{2^{-\nu'} \over \Gamma(1+\nu')}
   s^{\nu'},
\end{equation}
and
\begin{equation}
 \Gamma(2w)={2^{2w} \over 2\sqrt{\pi}}\Gamma(w)
 \Gamma\left(w+{1\over 2}\right),
\end{equation}
the asymptotic forms are explicitly written as
\begin{eqnarray}
 G^{+}_{(A)}(z',z'',\zeta)\rightarrow
 -{H^2\over 4\pi (2u)} &&
  {2\pi^{3/2}(\nu'+1) \over \sin\pi\nu' \sin2\pi\nu'
  \Gamma(\nu'+1) \Gamma(\nu'+3/2) [\Gamma(-\nu')]^2}
  e^{-(1+\nu')\zeta} (s's'')^{-(\nu'+1)}
\nonumber \\
  = {H^2\over 4\pi^{5/2}} &&
   \Gamma(-\nu'-1/2) \Gamma(\nu'+2) (2u)^{-\nu'-2},
 \\
 G^{+}_{(B)}(z',z'',\zeta)\rightarrow
 {H^2\over 4\pi (2u)} &&
 {\nu'  \Gamma(\nu'+1/2) \over \sqrt{\pi}\sin\pi\nu'
  \Gamma(\nu'+1)}
  e^{\nu'\zeta} (s's'')^{\nu'}
\nonumber \\
  = {H^2\over 4\pi^{5/2}} &&
   \Gamma(\nu'+1/2) \Gamma(-\nu'+1) (2u)^{\nu'-1}.
\end{eqnarray}
Comparing these with the asymptotic
form of the right hand side of Eq.~(\ref{deSinv}),
\begin{eqnarray}
 {H^2 \over 8\pi^2}{\Gamma(\nu'+2) \Gamma(-\nu'+1) \over
 \sqrt{u^2-1}}
&& \left(\alpha P^{-1}_{\nu'}(u)
 +{\beta\over\pi} \tan\pi\nu' Q^{-1}_{\nu'}(u)\right)
\nonumber \\
\rightarrow {H^2 \over 4\pi^{5/2}}
&& \left[
 \alpha\Gamma(\nu'+1/2) \Gamma(-\nu'+1)(2u)^{\nu'-1}
 +(\alpha+\beta)\Gamma(-\nu'-1/2) \Gamma(\nu'+2)(2u)^{-\nu'-2}
 \right],
 \label{asym}
\end{eqnarray}
it is easy to see that $G^{+}_{(A)}(z',z'',\zeta)$ corresponds to
the choice of the coefficients $\{\alpha=1,\beta=-1\}$,
and $G^{+}_{(B)}(z',z'',\zeta)$ to $\{\alpha=0,\beta=1\}$.
Thus the sum $\tilde G^{+}=G^{+}_{(A)}+G^{+}_{(B)}$
correctly gives the Euclidean vacuum Wightman function.
This completes our proof.

\section{Extension to $\nu'>0$}

Although we have assumed $-1/2\le\nu'<0$ or $\Re\nu'=-1/2$ to derive
the expression (\ref{Green}) for the Wightman function, if we examine
the proof in the previous section that it coincides with the Euclidean
vacuum Wightman function, we find it is valid irrespective of the
value of $\nu'$, that is $\tilde G^{+}=\tilde G^{+}_E$
 as long as the series converges.
Then if we consider the case $\nu'>0$, we find that
$\tilde G^{+}(z',z'',\zeta)$ differs from the original
$G^{+}(z',z'',\zeta)$, which is given by the
integral of products of the mode functions over real values of $p$,
Eq.~(\ref{Wightman}), or the contour integral (\ref{contourint}),
because different poles contribute to the integral.
Specifically the contribution of $a_0$ comes in and $c_0$ goes out
as $\nu'$ becomes positive.
Thus
\begin{eqnarray}
 \tilde G^{+}(z',z'',\zeta) - G^{+}(z',z'',\zeta)
 & = & -{H^2\over s' s''} \left(a_0 -c_0 \right)
 \nonumber \\
 & = &
 -{H^2\over 8\pi s' s'' \sinh\zeta}
  \Biggl[
  {\nu'\Gamma(2\nu'+1) \over \sin\pi\nu'} e^{-\nu'\zeta}
  P^{-\nu'}_{\nu'}(z') P^{-\nu'}_{\nu'}(z'')
\nonumber \\ && \hskip2cm
   -{\nu'\Gamma(2\nu'+1)\over\sin\pi\nu'}
   e^{\nu'\zeta}P^{-\nu'}_{\nu'}(z') P^{-\nu'}_{\nu'}(z'')\Biggr]
\nonumber \\
 & = &
 {H^2\over 4\pi^{5/2}} \Gamma (-\nu'+1) \Gamma (\nu'+1/2)
  {\sinh \nu'\zeta\over \sinh\zeta}(s's'')^{\nu'-1}.
 \label{diff}
\end{eqnarray}

Physical origin of this difference between $\nu'<0$ and $\nu'>0$
may be understood as follows.
The inverse transformation of Eq.~(\ref{oriG}) is given by
\begin{equation}
    \sum_{\sigma=\pm}
  \chi_{p,\sigma}(z')\overline{\chi_{p,\sigma}(z'')}
 ={4\pi a(t') a(t'') \over p}\int_{0}^{\infty}\sinh^2\zeta
 d\zeta\,{\sin p\zeta\over\sinh\zeta}\,G^{+}({t',0;t'',\zeta}).
 \label{inverse}
\end{equation}
As seen from the asymptotic behaviour of the Euclidean vacuum
Wightman function for large $\zeta$ given in Eq.~(\ref{asym}) with
$\{\alpha=1,\beta=0\}$,
the integral in Eq.~({\ref{inverse}) becomes manifestly divergent
for $\nu'>0$ if we set $G^{+}=G^{+}_E$.
This means that the Euclidean vacuum state
cannot be described by any power spectrum in the usual sense
in which the summation of products of the mode functions
is taken over real values of $p$.
In deriving the normalized set of mode functions in Section III,
we did not assumed anything other than that the Klein-Gordon norm
can be evaluated on two isolated hypersurfaces
denoted by (I) and (II).
Thus we conclude that the discrepancy at $\nu'>0$ is due to
the breakdown of this assumption.

Let us consider a linear functional space, $\cal F$,
which consists of all normalizable functions which satisfy
the field equation.
Here the normalizable functions mean that they have
finite Klein-Gordon norms, apart from the trivial factor of the
$\delta$-function.
The Klein-Gordon inner product is defined
on a Cauchy surface and is independent of the choice
of the surface.
To evaluate it, there is one convenient choice of the hypersurface,
(IV), which is specified by $r_C={\rm constant}$ and
included in the region $C$ as shown in Fig.~2.

As the system has the $O(3,1)$ symmetry, it is natural to
take the basis vectors of ${\cal F}$ from functions of
the separated form as given in Eq.~(\ref{sep}).
Restricting $p^2$ to be positive,
if we evaluate these mode functions on the hypersurface (IV),
they diverge at $t_C=\pi/2$ or $t_C=-\pi/2$ with infinite cycles of
oscillation.
However the oscillatory divergence is not so seriously bad
that the Klein-Gordon norm still has a finite value.
It then follows that two modes which have different
separation constants $p^2$ and ${p'}^2$ are orthogonal.
Therefore they are contained in ${\cal F}$ and may be chosen
to be the basis vectors.
Moreover, as the mode function $f_{pl}(r_C)$ decreases
rapidly as $r_C$ increases,
after the deformation of the hypersurface (IV) to that consists of
(I), (II) and (III), the contribution to
the Klein-Gordon norm from (III) can be neglected as was assumed
before. Thus the Klein-Gordon norm is correctly given by
the expression (\ref{KGip}).
For completeness, in Appendix A we show that the Klein-Gordon inner
products evaluated on the hypersurface (IV) are finite for mode
functions with $p^2>0$ and in fact coincide with those evaluated on
(I) and (II).

On the other hand, for $p^2<0$, almost all modes diverge at
 $t_C=\pi/2$ or $t_C=-\pi/2$ faster than $(t\mp\pi/2)^{-1}$
(the critical power exponent depends on the dimension)
and their Klein-Gordon norms diverge.
Hence they are not contained in ${\cal F}$.
However, when $\nu'>0$, there exists an exceptionally set of modes
with $p=i\nu'$ which behave well enough,
\begin{equation}
 u_{(*)lm} = H(-i\cos t_C)^{\nu'-1}{P^{-l-1/2}_{-\nu'-1/2}
 (i\sinh r_C)\over\sqrt{i\cosh r_C}} Y_{lm}(\Omega).
\end{equation}
Its Klein-Gordon norm is finite and is given by
\begin{eqnarray}
\langle u_{(*) lm},u_{(*) l'm'}\rangle
&&={i\cosh^2r_C\over H^2}\int_{-\pi/2}^{\pi/2}dt_C \cos t_C
  \int d\Omega
   \left\{{\partial u_{(*)lm}\over\partial r_C}\overline{u_{(*)l'm'}}
 -u_{(*)lm}{\partial\,\overline{u_{(*)l'm'}}\over \partial r_C}\right\}
 \nonumber \\
&&={2\sqrt{\pi}\,\Gamma(\nu')\over \Gamma(\nu'+1/2)
      \Gamma(-\nu'+l+1) \Gamma(\nu'+l+1)}\delta_{ll'}\delta_{mm'}
\nonumber\\
&&=:N_{(*)l}\delta_{ll'}\delta_{mm'}\,.
\end{eqnarray}
In the region $R$, these mode functions take the form,
\begin{equation}
u_{(*)lm}(x)=H(\sinh t_R)^{\nu'-1}{P^{-l-1/2}_{-\nu'-1/2}
(\cosh r_R)\over\sqrt{\sinh r_R}}Y_{lm}(\Omega).
\end{equation}
They have exactly the same functional form in the region $L$ as well,
with $(t_R,r_R)$ replaced by $(t_L,r_L)$.
Therefore the contribution to the Wightman function
from these modes in the region $R$ or $L$ becomes
(with the suffix $R$ or $L$ suppressed)
\begin{eqnarray}
 G^{+}_{(*)}(z',z'',\zeta)
 & := &\sum_{l,m}{u_{(*)lm}\overline{u_{(*)lm}}\over N_{(*)l}}
 \nonumber \\
&=&
{\Gamma(\nu'+1/2)\over2\sqrt{\pi}\, \Gamma(\nu')}
 \sum_{l,m}\Gamma(-\nu'+l+1) \Gamma(\nu'+l+1)
 \nonumber \\
&&\quad\times
  H^2 (s's'')^{\nu'-1}
{P^{-l-1/2}_{-\nu'-1/2}(\cosh r')\over \sqrt{\sinh r'}}
{P^{-l-1/2}_{-\nu'-1/2}(\cosh r'')\over \sqrt{\sinh r''}}
  Y_lm(\Omega') \overline{Y_lm(\Omega'')}
 \nonumber \\
& = &
 {H^2\over 8\pi^{3/2}}\,
\Gamma(-\nu'+1)\Gamma(\nu'+1/2)\,\nu'\,
 {P^{-1/2}_{-\nu'-1/2}(\cosh \zeta) \over \sqrt{\sinh \zeta}}
 \left[\lim_{r\rightarrow 0}
 {P^{-1/2}_{-\nu'-1/2}(\cosh r) \over \sqrt{\sinh r}}\right](s's'')^{\nu'-1}
 \nonumber \\
& = &
  {H^2\over 4\pi^{5/2}}\, \Gamma (-\nu'+1) \Gamma (\nu'+1/2)\,
  {\sinh \nu'\zeta\over \sinh\zeta}\,(s's'')^{\nu'-1},
 \label{imG}
\end{eqnarray}
where, in the third equality,
the relation required by the $O(3,1)$ invariance,
\begin{eqnarray}
 \sum_{l,m}4\pi\,
   {\Gamma(-\nu'+l+1) \Gamma(\nu'+l+1)\over
            \Gamma(-\nu'+1) \Gamma(\nu'+1)}
 &&{P^{-l-1/2}_{-\nu'-1/2}(\cosh r')\over \sqrt{\sinh r'}}
 {P^{-l-1/2}_{-\nu'-1/2}(\cosh r'')\over \sqrt{\sinh r''}}
  Y_lm(\Omega') \overline{Y_lm(\Omega'')}
\nonumber \\
 =&&
   {P^{-1/2}_{-\nu'-1/2}(\cosh \zeta) \over \sqrt{\sinh \zeta}}
   \lim_{r\to0}\left[
  {P^{-1/2}_{-\nu'-1/2}(\cosh r) \over \sqrt{\sinh r}}\right]
\nonumber\\
 =&& \sqrt{2\over\pi}\,
   {P^{-1/2}_{-\nu'-1/2}(\cosh \zeta) \over \sqrt{\sinh \zeta}},
\label{loren}
\end{eqnarray}
has been used. For $\nu'=-ip$ ($p>0$), this is just the equation
describing the completeness of $Y_{plm}$, Eq.~(\ref{completeness}),
which is well-known to hold.
In Appendix B, we show it holds for $\nu'>0$,
$\nu'\neq {\rm integer}$ as well.
Thus we find that
\begin{equation}
 \tilde G^{+}(z',z'',\zeta)
 = G^{+}(z',z'',\zeta) + G^{+}_{(*)}(z',z'',\zeta).
\end{equation}

As clear from the above, the degrees of
freedom described by $u_{(*) lm}$ cannot be quantized in the
open chart, where these modes are not normalizable.
However if the quantization is performed
on some other hypersurfaces, say,
the time constant hypersurfaces of a closed chart,
we do not confront with any problem in the quantization.
We may then quantize the field there, but instead of using the
harmonics on the three-sphere, we may expand the field operator
in terms of the mode functions associated with an open chart,
\begin{equation}
 \hat \phi(x)=\int_0^{\infty} dp \sum_{\sigma,l,m}
 v_{p\sigma lm}(x) \hat a_{p\sigma lm}
   + \sum_{l,m}v_{(*)lm} \hat a_{(*) lm} +h.c.;\quad 1>\nu'>0,
\end{equation}
where the orthonormalized mode functions are denoted by $v_\Lambda$ as before
(see Eq.~(\ref{Normmode}) for $\nu'<0$) and are now given by
\begin{eqnarray}
 v_{p\sigma lm}(x) &=&{H\over\sqrt{N_{p\sigma}}}
  {\chi_{p,\sigma}(t)\over\sinh t}f_{pl}(r)Y_{lm}(\Omega),
  \nonumber \\
 v_{(*)lm}(x) &=&{H\over \sqrt{N_{(*)l}}}(\sinh t)^{\nu'-1}
{P^{-l-1/2}_{-\nu'-1/2}(\cosh r)\over\sqrt{\sinh r}}Y_{lm}(\Omega).
\end{eqnarray}
In the above, the coordinates $(t,r)$ stand
for $(t_R,r_R)$ or $(t_L,r_L)$ if $x$ is in the region $R$ or $L$,
respectively, and for $(i(\pm t_C-\pi/2),r_C+i\pi/2))$
in the region $C$ where the upper (lower) sign corresponds to the
analytic continuation from the region $R$ ($L$) as specified in
Eq.~(\ref{connection}).

Thus we have found a complete prescription to express the Euclidean
vacuum state in terms of the mode functions associated with an open chart.
Even in general situations in which the background differs
from the exact de Sitter space and the mass of the scalar
field changes in time, the above method of decomposing the field
operator in terms of the mode functions which respect the $O(3,1)$
symmetry will be
a very useful tool to evaluate the evolution of the field as long as
the quantum state is initially in the Euclidean vacuum.
Furthermore, we expect that the logical procedure given here
should play an important role when we consider various problems
associated with the quantum tunneling through $O(4)$-symmetric
bubble nucleation.

\section{Conformal and minimal massless case}

In this section we consider the cases $\nu'=0$ and $\nu'\rightarrow1$,
which correspond to the conformal massless case and the minimal
massless limit, respectively.

\subsection{Massless conformal coupling case}

First we consider the case $\nu'=0$.
In this case the integration of Eqs.~({\ref{oriG}) and
({\ref{inverse}) can be easily performed
because the Legendre function reduces to
an elementary function as
\begin{equation}
 P_{0}^{ip} (\cosh t)={e^{-ip\eta}\over ip\Gamma(-ip)},
\end{equation}
where $\eta$ is related to $t$ by the relation,
 $\sinh\eta =-1/\sinh t$.
Then, the right hand side of Eq.~(\ref{oriG}) may be expressed as
\begin{equation}
{1\over 4\pi^2}\int_0^{\infty} dp {\sin p\zeta\over \sinh\zeta}
 \left\{{ 1 \over 1-e^{-2\pi p}} e^{-ip(\eta'-\eta'')}
 +{e^{-2\pi p}\over 1-e^{-2\pi p}} e^{ip(\eta'-\eta'')} \right\}.
 \label{eqthermal}
\end{equation}
This integral can be easily done by closing the contour and evaluating
its residues. Then we
find that it coincides with the Euclidean vacuum Wightman function,
\begin{equation}
 G^{+}(z',z'',\zeta)={H^2\over 8\pi^2}{1\over 1+u}.
 \label{confG}
\end{equation}

If we focus on either of the region $R$ or $L$, a natural choice of
the positive frequency function is
$\tilde\chi_p=e^{-ip\eta}/\sqrt{2p}$.
Interpreting Eq.~(\ref{eqthermal}) in terms of these mode functions,
the Euclidean vacuum is seen as a thermal state,
as was pointed out in \cite{Pfa},
This thermal nature may be understood as a consequence of
the loss of information.
For complete description of a quantum state, it is necessary to specify
the states both in the regions $R$ and $L$.
Hence if we consider the expectation value of an observable which has
support only in the region $R$ (or $L$),
it does not depend on the quantum state of the region $L$ (or $R$).
This effectively brings a pure state into a mixed state.
Of course, this phenomenon is not peculiar only for
$\nu'=0$ but it appears in the easiest form for this special case.
In fact, the expression for $\chi_{p\sigma}$ given in
Eq.(\ref{modefcn2}) does already indicate the thermal nature;
the ratio of the squared absolute values of the coefficients in front of
$P^{ip}_{\nu'}$ (``positive frequency'' in the natural sense)
and $P^{-ip}_{\nu'}$ (``negative frequency''), after normalized by
$N_{p\sigma}$ and summed over $\sigma=\pm$, is $e^{2\pi p}$.

\subsection{Massless minimal coupling limit}

Next let us consider the massless minimal coupling limit,
 $\nu'\rightarrow 1$.
Using the following formulae,
\begin{eqnarray}
 P^{-1}_{1+\epsilon}(u)
  &=&
  \left({u-1\over u+1}\right)^{1/2}\left\{
   {u+1 \over 2}+ {\epsilon \over 2}\left(
    {u-1 \over 2}+(u+1)\ln\left({u+1 \over 2}\right)\right)
    \right\} + O(\epsilon^2),
 \nonumber \\
 \Gamma(3+\epsilon)
  & =&
  2+\epsilon (3- 2\gamma) + O(\epsilon^2),
 \nonumber \\
 \Gamma(-\epsilon)
 & =&
 -{1\over\epsilon}-\gamma + O(\epsilon),
\end{eqnarray}
for $\epsilon\rightarrow0$,
the $\nu'\rightarrow 1$ limit of $G^{+}_E(z',z'',\zeta)$ is given by
\begin{equation}
G^{+}_E(z',z'',\zeta)\rightarrow
  {H^2 \over 8\pi^2}\left\{
  -{1\over \nu'-1}+\left[
  {1\over u+1}-\ln\left({u+1\over 2}\right) -2\right]
  +O(\nu'-1)\right\}.
 \label{nuone1}
\end{equation}
In the same limit, the contribution from $p=i\nu'$
modes to the Wightman function given by Eq.~(\ref{imG}) becomes
\begin{equation}
 G^{+}_{(*)}(z',z'',\zeta)
  \rightarrow-{H^2\over 8\pi^2 } \left\{
   {1\over \nu'-1}+\left[\zeta\coth\zeta
   +\ln\left({s's''\over 4}\right)+2\right]
   +O(\nu'-1) \right\}.
 \label{nuone2}
\end{equation}
On the other hand, for $\nu'=1$, Eq.~(\ref{oriG}) reduces to a very
simple form,
\begin{equation}
 a(\eta')a(\eta'')G^{+}(z',z'',\zeta)
 = {1\over 8\pi^2 }
  \int_{-\infty}^{\infty} dp{\sin p\zeta\over\sinh\zeta}
    {e^{\pi p}\over \sinh\pi p}
    {(z'+ip)(z''-ip) \over 1+p^2} e^{-ip(\eta'-\eta'')}.
\end{equation}
The above integration can be done elementarily to give
\begin{equation}
 G^{+}(z',z'',\zeta) ={H^2\over 8\pi^2}\Bigl\{
    {1\over 1+u}-\ln\left({1+u\over 2}\right)
    +\ln\left({s' s'' \over 4}\right)
      +\zeta\coth\zeta\Bigr\}.
 \label{nuone3}
\end{equation}
Adding Eqs.~(\ref{nuone2}) and (\ref{nuone3}), we immediately see
that $\tilde G^{+}(z',z'',\zeta)=G^{+}_E (z',z'',\zeta)$.
We note that the Wightman function same as Eq.~(\ref{nuone3})
has been obtained by Allen and Caldwell \cite{Bruce}
in a totally different fashion.

\section{Conclusion}

We have derived the expression for the Euclidean vacuum
Wightman function of a scalar field in de Sitter space
in terms of the mode functions associated with the harmonic functions
on an open chart, $Y_{plm}$, i.e., on hyperbolic time slices
which respect the $O(3,1)$ symmetry.
The formula we have obtained is applicable to any mass or curvature
coupling, provided the effective mass square is positive, i.e.,
$M^2_{eff}=M^2+12\xi H^2>0$.
Usually the power spectrum of the quantum fluctuations is described by
the modes with real values of comoving wavenumber $p$.
However we have found that the Euclidean vacuum Wightman function
cannot be described in this standard manner for $M^2_{eff}<2H^2$.
We have found that an additional contribution from the modes
with an imaginary value of $p$ is necessary to describe
the Euclidean vacuum Wightman function in this case.
 These modes are not square-integrable on a hyperbolic time slice
in an open chart but their Klein-Gordon norms are finite if
we take a complete Cauchy surface in de Sitter space to evaluate them.

The expressions for the Euclidean vacuum Wightman function and the
field operator we have obtained will be a
powerful tool to calculate observable quantities in open universe
inflation models such as the cosmic temperature fluctuations
when the initial quantum state is in the Euclidean
vacuum\cite{bucher,Bruce,sataya2}.
Furthermore, the technique developed in this paper to obtain a set of
orthonormalized mode functions associated with an open chart should be
applicable to more general situations provided the system has the
$O(3,1)$ symmetry. The issue of the quantum state after false vacuum
decay through $O(4)$-symmetric bubble nucleation is one interesting
example of such situations.
This issue has been discussed and investigated by several authors
\cite{rubakov,vacha,tasaya,styyPTP,tanaka} but in a rather formal
manner.
In particular, although a method to determine the unnormalized mode
functions after tunneling has been given,
 the normalization procedure applicable to general situations
has not been explicitly given, except for the case of flat spacetime
background\cite{yamamoto}.
Application of our normalization technique to some specific models of
false vacuum decay will be discussed in a forthcoming paper.

\acknowledgments

We would like to thank B. Allen, R. Caldwell, T. Hamazaki, D. Lyth,
and N. Turok for enlightening discussions. One of us (M.S.) would like
to thank for the hospitality given to him by the people at the Newton
Institute, Cambridge University, where a part of this work was done.
This work was supported in part by Monbusho Grant-in-Aid for
Scientific Research Nos. 2010, 2841 and 05640342 and the Sumitomo
Foundation.

\appendix

\section{}

Here we show the Klein-Gordon inner products on the hypersurface
 (IV) are finite for the mode functions with $p^2>0$ and they
coincides with those evaluated on the hypersurfaces (I) and (II).

We consider the mode function $u_{p\sigma lm}$ given by
\begin{equation}
u_{p\sigma lm}(x)={\chi_{p,\sigma}(t_C)\over a(t_C)}
                       f_{pl}(r_C)Y_{lm}(\Omega),
\end{equation}
where $a(t_C)=H^{-1}\cos t_C$,
$\chi_{p,\sigma}$ is given by Eq.~(\ref{modefcn}) and
$f_{pl}$ is by Eq.~(\ref{fpl}), both with relevant analytic
continuations to the region $C$ specified in Eq.~(\ref{connection}).
In particular, the analytic continuation of $f_{pl}$ to the region $C$
is given by
\begin{equation}
 f_{pl}(r_C) = {\Gamma(ip+l+1)\over\Gamma(ip)}
  {P^{-l-1/2}_{ip-1/2}(i\sinh r_C)\over \sqrt{i \cosh r_C}},
\end{equation}
which plays the role of the positive frequency function.
The Klein-Gordon inner products among $u_{p\sigma lm}$
 on the $r_C={\rm constant}$ hypersurface (IV) take the form,
\begin{eqnarray}
N_{p\sigma lm,p'\sigma' l'm'}
&:=& \Bigl\langle{ u_{p\sigma lm}\,,\,u_{p'\sigma' l'm'}}\Bigr\rangle
\nonumber\\
&=& {i\cosh^2 r_C\over H^2}
 \int_{-\pi/2}^{\pi/2} dt_C \cos t_C \int d\Omega
 \left\{{\partial u_{p\sigma lm} \over \partial r_C }
 \overline{u_{p'\sigma' l'm'}}
 -u_{p\sigma lm}{\partial\, \overline{u_{p'\sigma' l'm'}} \over
        \partial r_C}\right\}
 \nonumber \\
&=&N^{(1)}_{pl}N^{(2)}_{p\sigma,p'\sigma'}\delta_{ll'}\delta_{mm'}\,,
\end{eqnarray}
where
\begin{equation}
 N^{(1)}_{pl}
 := i\cosh^2 r_C
 \left\{{\partial f_{pl} \over \partial r_C }\overline{f_{pl}}
 -f_{pl}{\partial\overline{f_{pl}} \over \partial r_C }\right\},\quad
N^{(2)}_{p\sigma,p'\sigma'}
 :=\int_{-\pi/2}^{\pi/2} {dt_C\over\cos t_C}\,
\chi_{p,\sigma} \overline{\chi_{p',\sigma'}}\,.
\end{equation}
In the above, we have anticipated that the factor
$N^{(2)}_{p\sigma,p'\sigma'}$ should contain the delta function
$\delta(p-p')$ so that we may put
$p'=p$ when evaluating $N^{(1)}_{pl}$. This will be justified shortly.

The evaluation of the factor $N^{(1)}_{pl}$ is straightforward.
We immediately obtain
\begin{equation}
 N^{(1)}_{pl}={2p\over \pi}\,\sinh \pi p\,.
\label{None}
\end{equation}
The evaluation of the factor
$N^{(2)}_{p\sigma,p'\sigma'}$ needs a bit of consideration.
First, from the symmetry that
$\chi_{p,\sigma}(t_C)=\chi_{p,\sigma}(\sigma t_C)$ ($\sigma=\pm$),
we see that
$N^{(2)}_{p\sigma,p'\sigma'}$ must be proportional to
$\delta_{\sigma\sigma'}$.
Next, by using the field equation for $\chi_{p,\sigma}$,
\begin{equation}
-\left[{\partial\over\partial t_C} \cos^3 r_C
{\partial\over\partial t_C} {1\over \cos r_C}
+(\nu^2- {9\over 4})\cos^2 r_C
\right] \chi_{p,\sigma}=(1+p^2)\chi_{p,\sigma}\,,
\end{equation}
we find
\begin{equation}
(1+p^2)
\int_{-\pi/2}^{\pi/2} dt_C (\cos t_C)^{-1}
\chi_{p,\sigma} \overline{\chi_{p',\sigma'}}
=(1+{p'}^{2})
\int_{-\pi/2}^{\pi/2} dt_C (\cos t_C)^{-1}
\chi_{p,\sigma} \overline{\chi_{p',\sigma'}},
\end{equation}
which implies that $N^{(2)}_{p\sigma,p'\sigma'}=0$ if ${p'}^2\ne p^2$.
Hence it should be expressed as
\begin{equation}
N^{(2)}_{p\sigma,p'\sigma'}
=N^{(2)}_{p\sigma} \delta(p-p') \delta_{\sigma\sigma'}\,,
\end{equation}
where we have restricted $p$ and $p'$ to be positive (negative $p$ modes
are equivalent to positive ones because of the symmetry
$\chi_{-p,\sigma}=\chi_{p,\sigma}$).
Thus our remaining task is to evaluate $N^{(2)}_{p\sigma}$.

It turns out that $N^{(2)}_{p\sigma}$ for real $p$ can be evaluated
without detailed knowledge of the behavior of $\chi_{p,\sigma}$
 because the divergent contribution at $p'=p$ which
determines the coefficient of the $\delta$-function
comes only from the boundaries of integration at
$t_C=\pm\pi/2$.
Therefore, to obtain $N^{(2)}_{p\sigma}$, we only have to
know the behaviour of $\chi_{p,\sigma}$ near the boundaries.
For this purpose, we use the expression (\ref{modefcn2}) for
$\chi_{p,\sigma}$, which we recapitulate:
\begin{equation}
 \chi_{p,\sigma} =
 \left\{
 \begin{array}{l}
  \displaystyle
 {1\over 2\sinh\pi p}\left(
 {e^{\pi p}-\sigma e^{-i\pi\nu'}\over \Gamma(\nu'+ip +1)}
 P^{ip}_{\nu'}(z_R)-
 {e^{-\pi p}-\sigma e^{-i\pi\nu'}\over \Gamma(\nu'-ip +1)}
 P^{-ip}_{\nu'}(z_R)\right),
 \\\\
  \displaystyle
 {\sigma\over 2\sinh\pi p}\left(
 {e^{\pi p}-\sigma e^{-i\pi\nu'}\over \Gamma(\nu'+ip +1)}
 P^{ip}_{\nu'}(z_L)-
 {e^{-\pi p}-\sigma e^{-i\pi\nu'}\over \Gamma(\nu'-ip +1)}
 P^{-ip}_{\nu'}(z_L)\right).
 \end{array}
 \right.
\end{equation}
Then noting the asymptotic behavior of $P^{ip}_{\nu'}$
near the boundaries,
\begin{equation}
 P^{ip}_{\nu'}(z_R)\sim {2^{ip}t_R^{-ip}\over \Gamma(1-ip)}
 ={2^{ip}e^{-\pi p/2}(-t_C+\pi/2)^{-ip}\over \Gamma(1-ip)},
\quad
 P^{ip}_{\nu'}(z_L)\sim {2^{ip}t_L^{-ip}\over \Gamma(1-ip)}
 ={2^{ip}e^{-\pi p/2} (t_C+\pi/2)^{-ip}\over \Gamma(1-ip)},
\end{equation}
we find
\begin{equation}
 \chi_{p,\sigma}
 \sim
 \left\{
 \begin{array}{ll}
  \displaystyle
 \alpha_{p}(-t_C+\pi/2)^{-ip}-\beta_{p}(-t_C+\pi/2)^{ip}
 \quad &\hbox{\rm for}\quad t_C \rightarrow \pi/2-0,
 \\\\
  \displaystyle
 \sigma\left(\alpha_{p}(t_C+\pi/2)^{-ip}
          -\beta_{p}(t_C+\pi/2)^{ip}\right)
 \quad &\hbox{\rm for}\quad t_C \rightarrow -\pi/2+0,
 \end{array}
 \right.
\end{equation}
where
\begin{eqnarray}
 \alpha_{p} & = & {e^{-\pi p/2}(e^{\pi p}-\sigma e^{-i\pi\nu'})
    \over 2\sinh\pi p\,\Gamma(\nu'+ip +1)\Gamma(1-ip)},
 \nonumber \\
 \beta_{p} & = & {e^{\pi p/2}(e^{-\pi p}-\sigma e^{-i\pi\nu'})
    \over 2\sinh\pi p\,\Gamma(\nu'-ip +1)\Gamma(1+ip)}.
\end{eqnarray}
Then
\begin{eqnarray}
N^{(2)}_{p\sigma}\delta(p-p')
 =&&\lim_{\epsilon\to0}\biggl(
 -\int^0_{\epsilon}{dx\over x}
 \left\{\alpha_{p}\overline{\alpha_{p'}}x^{i(p'-p)}
        +\beta_{p}\overline{\beta_{p'}}x^{i(p-p')} \right\}
\nonumber\\
&&\qquad\quad
 +\int_0^{\epsilon}{dx\over x}
 \left\{\alpha_{p}\overline{\alpha_{p'}}x^{i(p'-p)}
        +\beta_{p}\overline{\beta_{p'}}x^{i(p-p')} \right\}\biggr)
\nonumber \\
=&& 2\pi\left\{
 \vert \alpha_{p} \vert^2
 +\vert \beta_{p} \vert^2 \right\} \delta(p-p'),
\end{eqnarray}
from which we obtain
\begin{equation}
N^{(2)}_{p\sigma}
 ={2\left(\cosh\pi p-\sigma\cos \nu'\pi\right)\over
   p\sinh\pi p\,\vert\Gamma(\nu'+ip+1)\vert^2}.
\label{Ntwo}
\end{equation}
{}From Eqs.~(\ref{None}) and (\ref{Ntwo}), we finally find
\begin{equation}
 N_{p\sigma lm,p'\sigma' l'm'} =
 {4\left(\cosh\pi p-\sigma\cos \nu'\pi\right)
 \over \pi\,\vert\Gamma(\nu'+ip+1)\vert^2}\,\delta(p-p')
 \delta_{\sigma\sigma'}\delta_{ll'}\delta_{mm'}\,.
\end{equation}
We see that this exactly coincides with the result obtained
 in Eq.~(\ref{chinorm}).

\section{}

In this Appendix we show the outline of a proof of
Eq.~(\ref{loren}). After summing over $m$
in the left hand side of the equation, the relation to be
proved becomes
\begin{eqnarray}
 \sum_{l=0}^\infty M_l
 &=&{P^{-1/2}_{-\nu'-1/2}(\cosh \zeta) \over \sqrt{\sinh \zeta}}
   \lim_{r\rightarrow 0}
   {P^{-1/2}_{-\nu'-1/2}(\cosh r)\over \sqrt{\sinh r}}
  =\sqrt{2\over\pi}
   {P^{-1/2}_{-\nu'-1/2}(\cosh \zeta) \over \sqrt{\sinh \zeta}},
\nonumber \\
 M_l&:=& {\Gamma(-\nu'+l+1) \Gamma(\nu'+l+1)
 \over \Gamma(-\nu'+1) \Gamma(\nu'+1)}
 {P^{-l-1/2}_{-\nu'-1/2}(\cosh r')
 \over \sqrt{\sinh r'}}
 {{P^{-l-1/2}_{-\nu'-1/2}(\cosh r'')}
 \over \sqrt{\sinh r''}}(2l+1) P_l(\cos\Theta),
 \label{tobeshown}
\end{eqnarray}
where
\begin{eqnarray}
 \cos\Theta &=& \cos\theta' \cos\theta'' +
\sin\theta' \sin\theta'' \cos(\phi'-\phi''),
\nonumber\\
 \cosh\zeta &=&\cosh r'\cosh r''-\sinh r'\sinh r''\cos\Theta,
\label{xidef}
\end{eqnarray}
and $P_l(\cos\Theta)$ is the Legendre function. Here we assume
$\nu'>0$ and $\nu'\neq{\rm integer}$.

The strategy is similar to the technique employed in Section III.
First we show that $\sum_l M_l$ converges. Then we show
the result depends only on $\zeta$.
Then taking the limit $r''\rightarrow0$, it is readily seen that
Eq.~(\ref{tobeshown}) holds.

First let us show the convergence of $\sum_l M_l$.
For this purpose, we rewrite $M_l$ as
\begin{equation}
 M_l=(2l+1){\Gamma(-\nu'+1)\Gamma(\nu'+l+1)\over
             \Gamma(\nu'+1)\Gamma(-\nu'+l+1)}
     F_l(r')F_l(r'')P_l(\cos\Theta),
\end{equation}
where the function $F_l(r)$ is defined by
\begin{eqnarray}
F_l(r)&:=&{\Gamma(-\nu'+l+1)\over\Gamma(-\nu'+1)}
         {P^{-l-1/2}_{-\nu'-1/2}(\cosh r)\over\sqrt{\sinh r}}
\nonumber\\
&=&\sqrt{2\over\pi}\,\Gamma(-\nu'+1)
    \sinh y\, e^{i\pi\nu'}Q^{-\nu'}_l(\cosh y),
\label{Fdef}
\end{eqnarray}
and $y$ is related to $r$ as $\sinh y=1/\sinh r$ or $y=\ln\coth r/2$.
With the aid of the integral representation,
\begin{eqnarray}
 e^{i\pi\nu'}Q^{-\nu'}_l(\cosh y)
 & =&{\sqrt{\pi}\sinh^{-\nu'} y\over \sqrt{2}
   \Gamma(\nu'+1/2)} \int^{\infty}_y
   {e^{-(l+1/2)x} dx \over (\cosh x -\cosh y)^{-\nu'+1/2}},
\nonumber\\
 &&
 \hbox{\rm for}\quad \Re\nu'>-1/2,
               \quad \Re(\nu'-l)<1,
               \quad y>0,
\end{eqnarray}
we see that $F_l(r)$ is real and positive for $l>\nu'-1$.
Further it forms a monotonically decreasing sequence for $l>\nu'-1$.
Hence it is bounded as
\begin{equation}
 0<F_l(r)<e^{-(l-n)y} F_n(r)=\left(\tanh{r\over2}\right)^{l-n}F_n(r),
\end{equation}
for $l>n=[\nu']$ where $[\nu']$ is the largest integer not exceeding
$\nu'$. Also for the same $n$,
\begin{equation}
 {\Gamma(\nu'+l+1)\over \Gamma(-\nu'+l+1)}
 <(\nu'+l)(\nu'+l-1)\cdots (\nu'+l-2n-1)<(\nu'+l)^{2n+2}.
\end{equation}
Thus, for sufficiently large $l$, the absolute value of $M_l$
is bounded from above as
\begin{equation}
|M_l|<{(2l+1)\Gamma(-\nu'+1)\over\Gamma(\nu'+1)}
(\nu'+l)^{2n+2}\left(\tanh{r'\over2}\tanh{r''\over2}\right)^{l-n}
 F_n(r')F_n(r''),
\end{equation}
which becomes exponentially small for large $l$.
Therefore the series $\sum_l M_l$ is manifestly convergent.
In particular, if $0<\nu'<1$, $F_n(r)$ is bounded from above for real
positive $r$ and approaches zero as $r\rightarrow\infty$.
Thus the series converges uniformly and absolutely for $0<\nu'<1$.
Furthermore, even for $\nu'>1$, as long as we keep $r$ finite, say
$r<r_{max}$, the series also converges uniformly and absolutely there.
In this sense, the convergence is uniform for any $\nu'>0$, $\nu'\neq
{\rm integer}$ (uniform convergence on compact sets).

Next we show the series $\sum_l M_l$ is indeed a function of
 $\zeta$ alone. The equation to be proved is
\begin{equation}
 \sum_l (\cos^2 \Theta -1)\left\{
 {\partial\zeta\over \partial\cosh r'}
 {\partial M_l\over\partial\cos\Theta}
 -{\partial\zeta\over\partial\cos \Theta}
 {\partial M_l\over\partial\cosh r'}\right\}=0.
\label{qqq}
\end{equation}
This calculation is a bit more complicated
than the previous case in Section III because the counterpart of
$e^{(\pm\nu-n) \zeta}$ there is $P_l(\cos\Theta)$ here.
Taking derivatives with respect to $\cos\Theta$ gives terms as
$\cos\Theta P_l(\cos\Theta)$ and $\cos^2 \Theta P_l(\cos\Theta)$.
The factor $\cos\Theta$ must be removed from these terms
by using the relation,
\begin{equation}
  \cos \Theta P_l(\cos \Theta)={1\over 2l+1}\left\{
   (l+1) P_{l+1}(\cos \Theta) +l  P_{l-1} (\cos \Theta)\right\}.
\end{equation}
After carrying out this procedure,
 the left hand side of Eq.~(\ref{qqq}) takes the form,
$\sum_l C_l (r',r'')P_l(\cos \Theta)$.
Then applying the recursion relation for the Associated Legendre
 functions, it can be shown that $C_l (r',r'')=0$.
 Thus the right hand side of Eq.~(\ref{tobeshown}) has the desired
invariance, which then immediately implies it is equivalent to the
 left hand side.


\medskip
\begin{center}
{\bf FIGURE CAPTIONS}
\end{center}
\medskip

\begin{itemize}
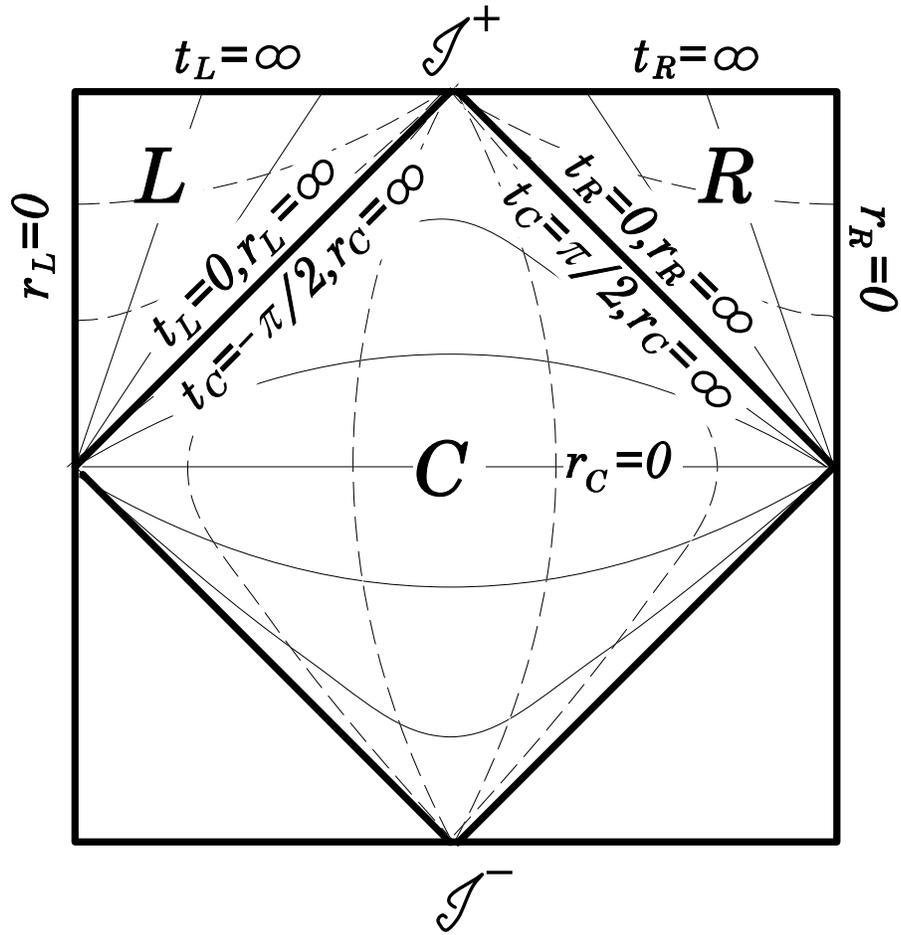
\item[Fig. 1.] Conformal diagram of de Sitter space with the
coordinates $(t,r)=(t_R,r_R)$, $(t_L,r_L)$ and $(t_C,r_C)$ spanning the
regions $R$, $L$ and $C$, respectively. The thin real lines denote
the surfaces $\{r={\rm constant}\}$ and the broken lines
$\{t={\rm constant}\}$.

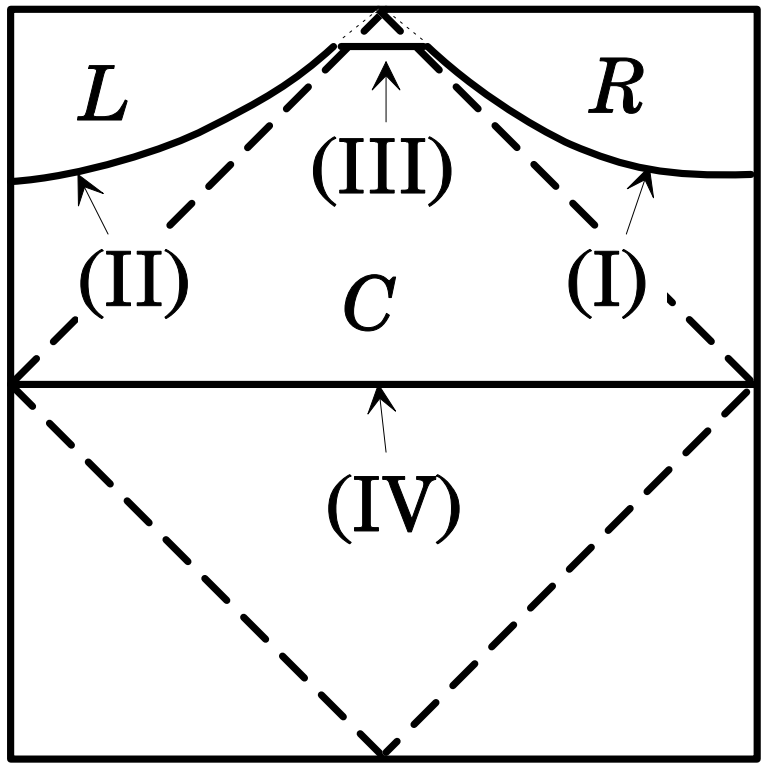
\item[Fig. 2.] Two different Cauchy surfaces on which the Klein-Gordon
innner products are evaluated. One of them consists of hyperbolic time
slices (I) and (II) in the regions $R$ and $L$, respectively,
 and the hypersurface (III) which connects them. The other one (IV) is
an $r_C={\rm constant}$ hypersurface entirely contained in the region $C$.

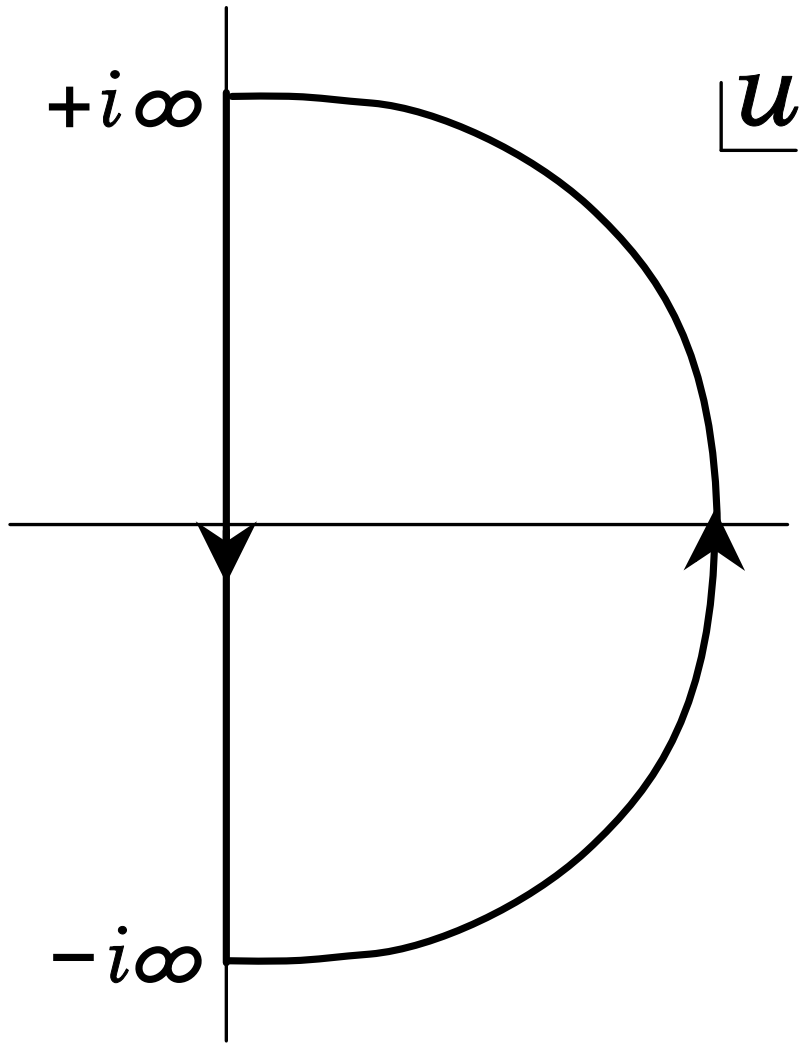
\item[Fig. 3.] The contour of integration on the complex $u$-plane for
Eq.~(\ref{contourint}).
\end{itemize}

\end{document}